\documentclass[aps,prb,reprint,amsmath,amssymb,superscriptaddress,notitlepage]{revtex4-2}
\usepackage[utf8]{inputenc}
\usepackage{bm}
\usepackage{enumerate}
\usepackage{graphicx}
\usepackage{xcolor}
\usepackage{comment}
\usepackage[T1]{fontenc}
\usepackage{booktabs}
\usepackage{hyperref}

\hypersetup{colorlinks=true,
linkcolor=blue,
citecolor=blue,
urlcolor=blue,
filecolor=blue
}
\usepackage{braket}

\newcommand{\kb}[2]{| #1 \rangle\langle #2 |}

\begin{document}

\title{Numerical simulation of coherent spin-shuttling in a {QuBus} with charged defects}

\author{Nils Ciroth}
\author{Arnau Sala}
\author{Ran Xue}
\affiliation{JARA-FIT Institute for Quantum Information, Forschungszentrum J\"ulich GmbH and RWTH Aachen University, Aachen, Germany}
\author{Lasse Ermoneit}
\author{Thomas Koprucki}
\author{Markus Kantner}
\affiliation{Weierstrass Institute for Applied Analysis and Stochastics (WIAS), Anton-Wilhelm-Amo-Str. 39, 10117 Berlin, Germany
}
\author{Lars R. Schreiber}
\email{lars.schreiber@physik.rwth-aachen.de}
\affiliation{JARA-FIT Institute for Quantum Information, Forschungszentrum J\"ulich GmbH and RWTH Aachen University, Aachen, Germany}
\affiliation{ARQUE Systems GmbH, 52074 Aachen, Germany}

\date{\today}

\begin{abstract}
Recent advances in coherent conveyor-mode spin qubit shuttling are paving the way for large-scale quantum computing platforms with qubit connectivity achieved by spin qubit shuttles. We developed a simulation tool to investigate numerically the impact of device imperfections on the spin-coherence of conveyor-mode shuttling in Si/SiGe. We simulate the quantum evolution of a mobile electron spin-qubit under the influence of sparse and singly charged point defects placed in the Si/SiGe heterostructure in close proximity to the shuttle lane. We consider different locations of a single charge defect with respect to the center of the shuttle lane, multiple orbital states of the electron in the shuttle with $g$-factor differences between the orbital levels, and orbital relaxation induced by electron-phonon interaction. With this simulation framework, we identify the critical defect density of charged point defects in the heterostructure for conveyor-mode spin qubit shuttle devices and quantify the impact of a single defect on the coherence of a qubit.
\end{abstract}

\maketitle

\section{Introduction\label{sec:intro}}
Spin qubits encoded into the spin of localized or mobile electrons or holes require both short- and long-range connectivity between qubits in order to fully harness the potential of quantum computation~\cite{DiVincenzo2000a,laucht_roadmap_2021,Vandersypen2017,burkard_semiconductor_2023}: Short range connectivity is necessary for implementing two-qubit gates between adjacent qubits, while long range connectivity is essential for coupling static or mobile distant qubits in scalable chip architectures~\cite{taylor_fault-tolerant_2005,Vandersypen2017, boter_spiderweb_2022,kunne_spinbus_2024} and for some implementations of surface codes that require interactions across the qubit lattice, such as toric codes~\cite{bravyi_quantum_1998}.

To address long-range connectivity, a variety of mechanisms have been proposed, including charge shuttling, spin-spin coupling, exchange-coupled spin chains, coupling via floating gates, superconducting resonators and cavities, dipole-dipole interactions, and super-exchange mechanisms~\cite{burkard_semiconductor_2023}. %
Shuttling of charges has been explored in GaAs by surface acoustic waves for charge~\cite{mcneil_-demand_2011,hermelin_electrons_2011} and spin~\cite{jadot_distant_2021,edlbauer_-flight_2021}, but the constant shuttle velocity, the generation of surface acoustic waves and lastly the large hyperfine interaction in GaAs limit the applicability to spin qubits. One promising method for qubit connectivity is spin-coherent conveyor-mode shuttling. In such a shuttle device, also called conveyor-belt QuBus~\cite{Seidler2022}, a single electron or hole is transported adiabatically across micrometer-scale distances by a periodic and moving confinement potential generated by electrostatic gates, called clavier gates~\cite{langrock_blueprint_2023}. The electrons or holes are confined to a moving quantum dot (QD) formed in a quantum well (QW). Due to the periodicity of the potential, only a few signals independent from the connection distance are required to control the voltage in the gates~\cite{xue_sisige_2024, Beer_2025}. It is compatible with fabrication in foundries~\cite{langheinrich_fabrication_2025} and has been successfully applied to non-piezoelectric qubit platforms such as electrons in tensile-strained Si/SiGe~\cite{struck_spin-epr-pair_2024, de_smet_high-fidelity_2024, matsumoto_two-qubit_2025}.
Its simplicity with respect to control signals triggered ideas towards shuttle-based scalable architectures~\cite{boter_spiderweb_2022, kunne_spinbus_2024, ginzel_scalable_2024, patomaki_pipeline_2024, siegel_snakes_2025} and is ideal for co-integration of cryo-electronics chips~\cite{zhao_ultra-low-power_2025}.

Despite successful demonstrations, a number of challenges persist mainly related to fabrication imperfections and disorder in the shuttle devices. Ge/SiGe suffer from variability of $g$-factor and spin-orbit interaction in conjunction with charge noise and strain~\cite{hendrickx_sweet-spot_2024, seidler_spatial_2025, rooney_gate_2025}. The SiGe alloy disorder in Si/SiGe has an impact on valley splitting and valley excitation of electrons~\cite{paquelet_wuetz_atomic_2022, klos_atomistic_2024, losert_practical_2023, thayil_2025}. This issue has been in focus of material mapping~\cite{volmer_mapping_2023, volmer_reduction_2025} and mitigation strategies~\cite{losert_strategies_2024, david_long_2024}. A common challenge is the electrostatic disorder in a shuttle device, which cannot be easily tuned away if only a few control signals shall be used~\cite{klos_calculation_2018}. In the worst case, this disorder can locally jeopardize the charge confinement of the moving QD with loss of qubit position~\cite{xue_sisige_2024}, but it might also lead to uncontrolled orbital and valley excitations/relaxations within the moving QD with impact on the spin-coherence~\cite{langrock_blueprint_2023, volmer_reduction_2025}. The geometry of shuttle devices is usually optimized to minimize the impact of disorder from remote charge defects at the semiconductor-oxide interface~\cite{langrock_blueprint_2023}, line-etch roughness of gates~\cite{seidler_tailoring_2023}, and strain effects induced by the metallic gates~\cite{corley-wiciak_lattice_2023}. Sparse defects proximal to the moving QDs, however, might have a significant impact. Among sparse defects are charged point defects---originating from unintentional dopants or fabrication-induced imperfections---, threading dislocations from the relaxed SiGe virtual substrate, and misfit-dislocations at the Si/SiGe interface~\cite{liu_role_2022}. Previous numerical studies have explored various aspects of electron shuttling dynamics~\cite{buonacorsi_simulated_2020,jeon_robustness_2024}, providing a qualitative insight into the physical interactions that can affect the shuttling operation.

Our work focuses on simulating the time-dependent shuttling of a spin qubit in a QuBus in the presence of discrete, localized charged defects, while explicitly accounting for the complexity of spin dynamics. We simulate the full quantum time evolution of the electron's envelope wave function using realistic electrostatic potentials and include multiple orbital states, spin dynamics, $g$-factor inhomogeneities across orbital states, and orbital relaxation processes due to electron-phonon interactions.
This enables us to assess how individual charged defects affect the shuttling fidelity and spin coherence in a QuBus with sparse defects and under different modes of operation.

Our results shed light on the operational limits of the QuBus architecture and provide guidance for optimizing its design and operation in the presence of negative single charge defects in the bulk Si channel.
By providing a more realistic and time-resolved model of defect-induced decoherence during spin shuttling, our work complements and extends previous theoretical and experimental studies~\cite{jeon_robustness_2024,losert_practical_2023}, and contributes toward establishing conveyor-mode spin shuttling as a robust mechanism for scalable quantum computing with silicon spin qubits.

The paper is organized as follows: In Sec.~\ref{sec:model} we introduce the device under consideration, its mode of operation, the expected forms of disorder, and the electrostatic model. In Sec.~\ref{sec:sim} we introduce the Hamiltonian and the theoretical framework used for the time-dependent simulations of the spin dynamics. The corresponding results are presented in Sec.~\ref{sec:results}. We conclude in Sec.~\ref{sec:conclusions} with a discussion on the scaling of the dephasing metrics in larger devices and with varying defect densities. Additional details on the model and the numerical simulations are described in the Appendices~\ref{app:comsol}-\ref{app:methods}.

\begin{figure}
    \centering
    \includegraphics[width=\linewidth]{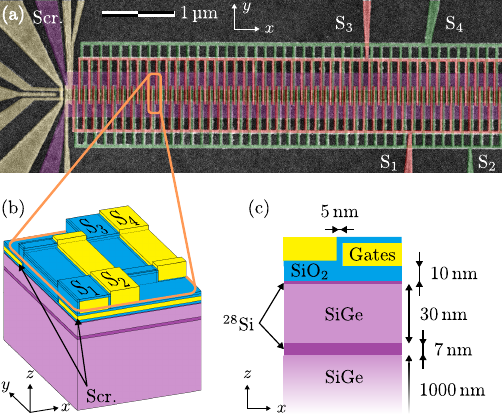}
    \caption{(a)~Scanning electron micrograph of one of the QuBus devices that we model. This device has a set of shuttling gates shown in red and green (labeled S$_1$--S$_4)$, two screening gates (Scr.) in purple, and individually controllable clavier gates and a single-electron transistor in yellow. We consider a numerical model of the unit cell (b) contained in the orange rectangle (a) and impose periodic boundary conditions along the channel ($x$-direction), see Appendix~\ref{app:comsol} for details. (c)~The heterostructure and gate stack consists of a large SiGe substrate upon which an isotopically purified \textsuperscript{28}Si layer with a thickness of 7\,nm is grown. This is the electron channel, where we can accumulate a 2DEG and shuttle the electrons. On top of that, we consider another SiGe buffer and a thin Si cap layer.  A SiO$_2$ layer separates the semiconductor from the metallic gates to avoid Schottky contacts.}
    \label{fig:qubus}
\end{figure}

\section{Model\label{sec:model}}
\subsection{Conveyor-mode shuttling in the QuBus\label{sec:CM}}
\begin{figure}
    \centering
    \includegraphics[width=\linewidth]{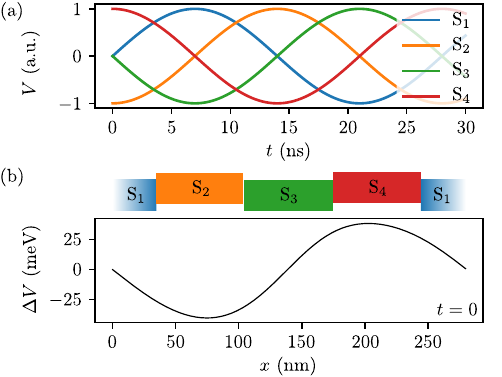}
    \caption{(a) Voltages applied to the shuttling gates and simulated potential. Each voltage $U_i(t)$ may have a different offset $U_{i}^\text{DC}$ and amplitude $U_{i}^\text{AC}$. For simplicity, we used the same offset and amplitudes for all the gates in this case. (b) Calculated potential in the shuttling channel at a fixed time $t=0$. Here, we applied $U_\text{S$_1$}^\text{DC} = U_\text{S$_3$}^\text{DC} = 550\,$mV and $U_\text{S$_2$}^\text{DC} = U_\text{S$_4$}^\text{DC} = 700\,$mV, with a pulse amplitude $U_\text{S$_1$}^\text{AC} = U_\text{S$_2$}^\text{AC} = U_\text{S$_3$}^\text{AC} = U_\text{S$_4$}^\text{AC} = 280\,$mV, which results in a fairly sinusoidal spatial confinement along the shuttling direction.}
    \label{fig:clav_gates}
\end{figure}

Conveyor-mode shuttling is the ideally adiabatic transport of an electron confined in a moving gate-defined QD~\cite{langrock_blueprint_2023}. This can be achieved in a two-dimensional electron gas (2DEG) channel with an array of clavier gates on top, perpendicular to the channel, as shown in the QuBus device of Fig.~\ref{fig:qubus}. In this geometry, four sets of repeating gates (S$_1$, S$_2$, S$_3$ and S$_4$) on top of a 2DEG channel modulate the potential landscape along the $x$-direction, while two screening gates (Scr., in purple) confine the shuttling channel along the $y$-direction. With four independent clavier gates one can form a QD, by applying time-dependent potentials $U_\text{S$_1$}$, $U_\text{S$_2$}$, $U_\text{S$_3$}$ and $U_\text{S$_4$}$ on the respective gates, as depicted in Fig.~\ref{fig:clav_gates}a. Applying this configuration of voltages to all the gates in the channel with a four-fold periodicity results in an array of QDs formed by an approximately sinusoidal potential in the shuttling channel (Fig.~\ref{fig:clav_gates}b), where the valleys of the potential confine an electron and the peaks are barriers between the QDs. %
A timely sinusoidal modulation of these potentials, as in
\begin{align}\label{eq:shuttle}
    U_i (t) ={}&{} U_{i}^\text{DC} + U_{i}^\text{AC}\sin\left(2\pi \nu t - \phi_i\right),
\end{align}
with a frequency $\nu$ moves the QDs with a velocity $v=\nu\lambda$, where $\lambda/4$ is the gate pitch. The voltages at neighboring gates are phase-shifted by $\phi_{i+1}- \phi_i = \pi/2$.
The DC components $U_{i}^\text{DC}$ can be adjusted to account for the different vertical distances of clavier gates from the shuttling channel, see Fig.~\ref{fig:qubus}b, and
the drive amplitude $U_{i}^\text{AC}$ controls the confinement depth and the barrier height between neighboring QDs. Ideally, with a gate pitch of $\lambda/4=70\,$nm, and a frequency $\nu=35.7$\,MHz, an electron confined to one of these QDs is thus shuttled adiabatically at a velocity $v=10\,$m/s, akin to a conveyor belt.

\subsection{Disorder in the QuBus\label{sec:disorder}}
During conveyor-mode shuttling, an electron interacts with its environment, which can affect the coherence or even change its quantum state in a measurable way. This effect has been leveraged to probe material parameters such as the local valley splitting~\cite{volmer_mapping_2023} or the electrostatic disorder~\cite{xue_sisige_2024} within the channel. As such, conveyor-mode shuttling via a QuBus requires high quality devices with very few defects, as any form of noise or inhomogeneities of the electrostatic landscape can interfere with the qubit and introduce a source of decoherence.

Due to the weak intervalley relaxation mechanism~\cite{penthorn_direct_2020}, valley excitations can only impact the dephasing of spin qubits if there is a strong enough spin-valley interaction or a $g$-factor difference across valley states. Otherwise, the spin of the electron will remain unchanged during the evolution, irrespective of intervalley transitions during shuttling. %
In contrast, although the spin-orbit interaction in Si/SiGe heterostructures is weak, (non-adiabatic) orbital excitations can lead to substantial spin dephasing as a result of small $g$-factor differences between orbital states.

If we neglect the effects of valley excitations, then %
a major mechanism of decoherence consists of an excitation of a higher orbital due to electrostatic disorder induced by a charged defect in the channel. While the density of these defects is small in a state-of-the-art heterostructure, the probability of interacting with one during coherent shuttling is not negligible. Assuming a density of single-negatively charged defects $n_d\sim 10^{14}$\,cm$^{-3}$ and a uniform distribution of defects throughout the semiconductor at 100\,mK, the probability of interacting with at least one such defect during a typical shuttling operation follows a
 Poisson distribution $P_{\geq1} = 1-\exp{\left(-\lambda\right)}$, where $\lambda=n_d V$. Here $V$ is the volume swept by the shuttled electron, given by $V=L\times W\times H$, where $L$ is the shuttling length and $W$ and $H$ are the effective width and height or interaction length along the $y$ and $z$ direction, respectively. 
For a typical QuBus with a length $L=10$\,µm, the expected cross-section of interaction is in an interesting regime: Assuming vertical and lateral interaction lengths roughly corresponding to the span of the envelope wave function of the electron, with $W\approx 20$\,nm and $H\approx 3$\,nm, the resulting probability is of the order of 1\% to 5\%. In the case of a shallow QD with weak confinement, where the envelope wave function spans over a larger space, or a longer shuttling distance, this probability can even go up to 20\%. This large probability and its uncertain range motivate the study of the coherence of a shuttled electron in a QuBus with sparse charge defects, both to pin down the probabilities further and to investigate the impact of such an interaction with a defect.

The limited set of interactions under consideration makes our discussion relevant for Si/SiGe devices, but also for Ge/SiGe devices and some monolayered materials such as transition metal dichalcogenides, where either there are no valley states or the valley degeneracy has been lifted by a strong spin-orbit interaction.

\subsection{Modeling the electrostatic potential\label{sec:potential}}

We model the electrostatic potential landscape in the QW of the QuBus induced by the electrostatic gates, and a single-point charge defect with the negative elementary charge $e$. The electrostatic calculations are carried out by solving Poisson's equation using COMSOL Multiphysics\textsuperscript{\textregistered}. We assume Dirichlet boundary conditions at the gate electrodes and periodic boundary conditions along the shuttling axis, see Fig.~\ref{fig:qubus}b-c. %
Due to the strong confinement in the vertical direction, the electron's envelope wave function effectively factorizes into a vertical and a two-dimensional in-plane component. In order to simulate the orbital dynamics within the 2D plane, we extract a 2D cross-section of the full electrostatic potential at the center of the QW.

We assume negligible nonlinear effects arising from charged regions near the metal-oxide contacts, and exploit the resulting linearity of Poisson's equation to separate the problem into two steps: (1)~We evaluate the potential generated by the charge defect $V_\text{def}$, modeled as a uniform charge distribution of total charge $e$ occupying one element of the tetrahedral mesh, and (2)~we evaluate the potential generated by the electrostatic gates $V_\text{gates}(t)$ at a given time $t$. Finally, we reassemble the two potentials as $V(t) = V_\text{def} + V_\text{gates}(t)$. We evaluate $V_\text{def}$ for different positions of the defect, and $V(t)$ for each time $t$. Separating these two potentials and then recombining them when needed drastically reduces the complexity of the problem. Appendix~\ref{app:comsol} contains more details about the electrostatic model.

\section{Time-dependent simulations\label{sec:sim}}
\subsection{Hamiltonian and the adiabatic framework\label{sec:ham}}

Using the numerically computed time-dependent potential $V(t)$, we solve the time-dependent Schrödinger equation for the envelope wave function of the electron using the adiabatic frame expansion. We start by considering a real-valued, time-dependent Hamiltonian $H(t)$, with instantaneous eigenstates $\ket{\phi_{l,\sigma}(t)}$ and eigenenergies $E_{l,\sigma}(t)$
\begin{equation}\label{eq:instantaneous-eigenstates}
    H(t)\ket{\phi_{l,\sigma}(t)} = E_{l,\sigma}(t)\ket{\phi_{l,\sigma}(t)},
\end{equation}
where $l$ labels the orbital quantum number and $\sigma$ is the spin state.
The Hamiltonian at hand contains an orbital part with the simulated electrostatic potential $V(t)$:
\begin{align}\label{eq:orbital-Hamiltonian}
    H_\text{o}(t)
    =
    -\frac{\hbar^2}{2}\nabla\cdot \left(m^{-1}\nabla\right)
    +
    V(t),
\end{align}
where $m$ is the effective mass of the electron in the Si QW \cite{Schaeffler1997}, and a Zeeman Hamiltonian of the form
\begin{align}\label{eq:Zeeman}
    H_Z(t)
    &{
    =    
    \sum_{l,\sigma} \frac{g_l \mu_B}{\hbar} B \hat{S} \kb{\phi_{l,\sigma}(t)}{\phi_{l,\sigma}(t)}
    }
\end{align}
where $\mu_B$ is the Bohr magneton, $B$ is the magnetic field along the spin quantization axis and $\hat{S}$ is the corresponding spin projection operator acting as $\hat{S}\ket{\phi_{l,\sigma}} =  \hbar \sigma \ket{\phi_{l,\sigma}}$ for $\sigma=\pm1/2$. Note that we consider a magnetic field oriented perpendicular to the QW growth axis in the present device.
The $g$-factors $g_l$ depend on the orbital quantum number of the wave function of the electron. Experimental studies have shown a weak but significant dependence on the orbital level (see e.g.,~\cite{kawakami_electrical_2014,corna_electrically_2018}) that we approximate as $(g_l - g_0)/g_0\simeq 10^{-3}$ for $l>0$, where we use $g_0=2$ for the $g$-factor of an electron in the ground state of a silicon QD.

\begin{figure}
    \centering
    \includegraphics[width=\linewidth]{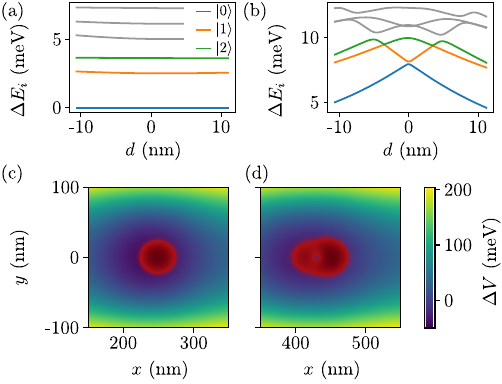}
    \caption{Influence of the defect on the energy spectrum and the wave function of the shuttled electron. Energy spectra of the minimum of the QD (a) without and (b) with a defect as a function of traveled distance ($d=vt$). In (b) the defect is localized at $d=0$\,nm, and in both cases we consider a magnetic field $B=0$\,T and plot the energy levels of one spin component only. (c-d) Envelope wave function of the shuttled electron (red) in a QD (c) away from the defect and (d) at the defect position. In all panels, the confinement potential is the same as in Fig.~\ref{fig:clav_gates}.}
    \label{fig:spectra}
\end{figure}

Diagonalization of the time-dependent Hamiltonian $H(t) = H_\text{o}(t) + H_Z(t)$ via a finite-difference method yields a locally flat energy spectrum when the electron is far from the defect, as shown in Fig.~\ref{fig:spectra}a. Here we show the six lowest energy levels, relative to the ground state. The energy difference between the ground state $\ket{0}$---an $s$ orbital---and the next excited state is quite constant along this short distance, with only minor variation of the energy splitting due to the staggered arrangement of the clavier gates. The finite energy splitting between the next two orbitals $\ket{1}$ and $\ket{2}$, corresponding to $p_x$ and $p_y$ orbitals, respectively, shows that the QDs are not perfectly circularly symmetric. The other energy levels, in gray, correspond to the in-plane $d$ orbitals. These have only a minor influence on the dynamics of the electron near the position of the defect. 
Due to these large energy splittings and the smoothness of the potential, for most of the quantum evolution simulation only three orbital levels are necessary to simulate the evolution of the envelope wave function of the electron with a reasonable accuracy. Near the position of the defect, energy anticrossings appear (see Fig.~\ref{fig:spectra}b), and their corresponding hybridization of states induce Landau--Zener type transitions to excited orbital states. For the considered shuttling velocity, we observe fully converged results when considering six orbital states.

After diagonalizing the Hamiltonian $H(t)$ at each time step, a general form of the wave function of the shuttled electron can be expressed in the basis of the instantaneous eigenstates $\{\ket{n(t)}\}$ as
\begin{align}\label{eq:Schr}
    \ket{\psi(t)} = \sum_n \tilde c_n (t) \ket{n(t)},
\end{align}
where the multi-index $n$ labels pairs of orbital and spin states $(l_n, \sigma_n)$. In the case of a real Hamiltonian 
with a smooth potential, 
any geometric phase will be 0 (mod $2\pi$), such that the fast evolution of the amplitudes $\tilde{c}_{n}$ can be eliminated by separating out the dynamical phases $\varphi_n (t) = \frac{1}{\hbar} \int_0^t \mathrm{d}t'\, E_n(t')$ as $\tilde c_n(t) = c_n (t) e^{-i\varphi_n (t)}$.
Substituting the expansion~(\ref{eq:Schr}) in the time-dependent Schrödinger equation yields an evolution equation for the complex-valued, time-dependent coefficients $c_n (t)$.
Using the Hellmann--Feynman theorem \cite{cohen-tannoudji_1977, Sakurai_Napolitano_2020}, the relevant equation of motion is obtained as
\begin{align}\label{eq:eq_motion}
    \dot{c}_n (t) ={}&{} \sum_{m\neq n}  \frac{\bra{n(t)}\dot{H}(t) \ket{m(t)}}{E_n(t) - E_m(t)} e^{-i\varphi_{mn} (t)}
    c_m (t),
\end{align}
where $\varphi_{mn}(t)=\varphi_{m}(t)-\varphi_{n}(t)$.

\subsection{Orbital relaxation and decoherence}
We introduce orbital relaxation into the system by introducing electron-phonon coupling to our model. We consider only spin-conserving transitions, and neglect the effects of the crystallographic and valley structure on the wave function of the electron~\cite{Saraiva2011,Tahan2014}. 
For this we use the Hamiltonian for electron-phonon interaction~\cite{Raith2011,Tahan2014}
\begin{align}\label{eq:ham_eph}
    H_\text{e-ph} ={}&{} i \sum_{\mathbf{k},p} \sqrt{\frac{\hbar k}{2\rho_0 \mathcal{V} v_p}} \Xi^{(p)} \left[ a_{\mathbf{k},p}^\dagger e^{i \mathbf{k}\cdot \mathbf{r}} - a_{\mathbf{k},p} e^{-i \mathbf{k}\cdot \mathbf{r}} \right].
\end{align}
Here, $a_{\mathbf{k},p}^\dagger$ and $a_{\mathbf{k},p}$ are creation and annihilation operators acting on the phonon space, i.e., the creation operator acting on the vacuum, $a_{\mathbf{k},p}^\dagger \ket{\text{vac}} = \ket{1_{\mathbf{k},p}}$, creates a phonon with momentum $\mathbf{k}$ and polarization $p$. The polarization-dependent deformation potential $\Xi^{(p)}$ considered here encapsulates the contribution from the uniaxial shear and the dilation deformation as $\Xi^{(p)} = \Xi_d \mathbf{e}_k^{(p)}\cdot \mathbf{k} + \Xi_u (\mathbf{e}_k^{(p)})_zk_z$, where $\mathbf{e}_k^{(p)}$ are the unit vectors along the directions $p$ of the phonon polarization. In this expression, we have also used the mass density of the crystal $\rho_0$, the crystal volume $\mathcal{V}$ and the phonon velocity $v_p$, which is different for each polarization direction $p = \{l,t_1,t_2\}$. Using the set of parameters from Ref.~\cite{Raith2011}, and the envelope wave functions and energies evaluated at each time step of the simulation, we evaluate the instantaneous relaxation rate for an electron from a state $\ket{n(t)}$ to a state $\ket{m(t)}$ with the emission of a phonon using Fermi's golden rule
\begin{align}\label{eq:fgr}
    \Gamma_{mn}(t) ={}&{} \sum_f \frac{2\pi}{\hbar} \left| \bra{m(t);1_{\mathbf{k},p}} H_\text{e-ph} \ket{n(t);\text{vac}} \right|^2 \notag \\
    {}&{}\times \delta(E_m(t) + \hbar\omega_{k,p} - E_n(t)),
\end{align}
where the sum runs over all final states (phonon momenta and polarization), and the delta function ensures energy conservation. In this expression $\hbar\omega_{k,p}$ is the energy of a photon with momentum $\mathbf{k}$ and polarization $p$, assuming the shuttling axis is aligned with the [100] crystallographic direction.

We derive a Lindblad equation for the system's density matrix by following the same steps employed in the derivation of Eq.~(\ref{eq:eq_motion}). By expanding the density matrix as
\begin{align}
    \rho(t) ={}&{} \sum_{n,m} r_{mn} (t) e^{-i \varphi_{mn} (t)} \kb{m(t)}{n(t)},
\end{align}
we obtain an equation of motion of the coefficients $r_{mn}$ of the density matrix
\begin{widetext}
    \begin{align}\label{eq:lindblad}
    \dot{{r}}_{kl} (t) ={}&{}
    \sum_{m\neq k} \frac{\bra{k(t)}\dot{H}(t) \ket{m(t)}}{E_k(t) - E_m(t)} e^{-i \varphi_{mk}} {r}_{ml}(t)
    - \sum_{n\neq l}  \frac{\bra{n(t)}\dot{H}(t) \ket{l(t)}}{E_n(t) - E_l(t)} e^{-i \varphi_{ln} (t)} {r}_{kn}(t)\notag \\
    {}&{} + \sum_{j,m,n} \Gamma_j  e^{-i[\varphi_{mn} (t) - \varphi_{kl} (t)]} \bra{k(t) }\left\{ C_j \kb{m(t)}{n(t)} C_j^\dagger 
    - \frac{1}{2} C_j^\dagger C_j \kb{m(t)}{n(t)}  - \frac{1}{2} \kb{m(t)}{n(t)} C_j^\dagger C_j \right\} \ket{l(t)}r_{mn}(t),
    \end{align}
\end{widetext}
where the collapse operator $C_j$ brings the electron from a state $s$ to a lower energy state $r$ with $E_s > E_r$, and a relaxation rate $\Gamma_j$. We label each allowed transition with the pair $(s_j, r_j)$ so that we can compactly write collapse operators and the relaxation rates as $C_j$ and $\Gamma_j$, respectively.
The time evolution of the density matrix of the electron is computed numerically via a second-order Magnus expansion scheme as described in Appendix~\ref{app:methods}. In order to characterize the coherence of the electron during shuttling in the presence of a charged defect, we assume that the electron is initially in the orbital ground state with an equal superposition of spin states, i.e.,
\begin{align}
    \ket{\psi(t=0)} = \frac{1}{2}\ket{0} \otimes \left( \ket{\uparrow} + \ket{\downarrow} \right).
\end{align}

The numerical results show that far away from the defect, the electron wave function is nearly circularly symmetric, as shown in Fig.~\ref{fig:spectra}c. As the electron approaches the defect, the defect potential changes the character of the envelope wave function. At the position of the defect (Fig.~\ref{fig:spectra}d), the wave function of the electron resembles a $p$ orbital, with a node at the location of the defect. At this point, the orbitals hybridize. Here, we observe a Landau--Zener transition that partially brings the electron to an excited orbital state.

During the evolution of the density matrix, we monitor a few figures of merit to reconstruct observables that can quantify the decoherence of the electron as it is shuttled across a defect.

\section{Results \label{sec:results}}
\begin{figure}
    \centering
    \includegraphics[width=\linewidth]{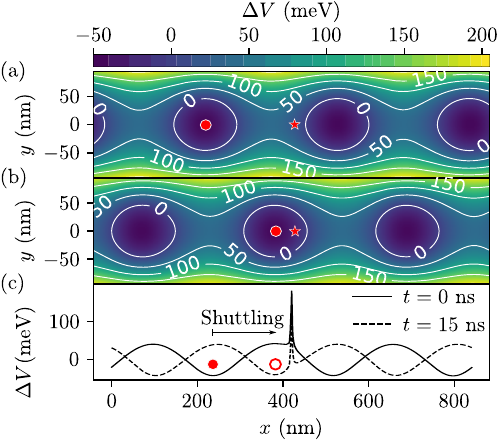}
    \caption{Numerically computed two-dimensional potential at the center of the QW at a fixed time. The sinusoidal potential applied to the S$_i$ gates combined with the effect of the screening gates results in a periodic array of QDs. (a)~At $t=0$\,ns the electron (red circle) is confined in one of the dots, away from the charge defect (red star). (b)~Via time-dependent modulation of the confinement potential, the electron is shuttled towards the defect. (c)~Longitudinal cross-section of the potential landscape. The potential barrier induced by the point-like defect exhibits a $1/r$ singularity at the defect position. The red full circle indicates the electron's initial position ($t=0$\,ns) and the empty one shows the electron's position after being shuttled at $t=15$\,ns. 
    }
    \label{fig:potential}
\end{figure}

Throughout the quantum evolution simulations, we have used a voltage on the screening gates $U_\text{Scr}^\text{DC} = 0$~V, and voltages $U_\text{S$_1$}^\text{DC}=U_\text{S$_3$}^\text{DC}=550$~mV and $U_\text{S$_2$}^\text{DC}=U_\text{S$_4$}^\text{DC}=700$~mV on the clavier gates, which results in the fairly harmonic confinement potential of Fig.~\ref{fig:potential}. The difference between these voltages is necessary due to the position difference of the gates along the vertical direction. The voltages are sinusoidally modulated, with a driving amplitude $U_{i}^\text{AC} = 280$~mV, equal for all the shuttling gates and a frequency $\nu = 35.7$~MHz. At the initial time, an electron---red dot in Fig.~\ref{fig:potential}a--- is assumed to be in the orbital ground state of a chosen QD. As the potential evolves as Eq.~(\ref{eq:shuttle}), the QD moves towards a point-like defect, marked as a red star in Figs.~\ref{fig:potential}a-b. Due to the large orbital energy splitting of about 3 meV far away from the defect, see Fig.~\ref{fig:spectra}a, the shuttling is nearly perfectly adiabatic for this velocity.
The charged defect poses a large potential obstacle right in the middle of the channel, see Fig.~\ref{fig:potential}c.
When the dot is swept over the defect, the QD is effectively split into a double-dot, which induces a sudden drop in the orbital energy splitting and strong hybridization of (instantaneous) orbital states. This gives rise to non-adiabatic time evolution, i.e., population of excited orbital electronic states (Landau--Zener type excitation). Due to $g$-factor differences among orbital states, this defect-induced process leads to spin-dephasing.

\begin{figure}
    \centering
    \includegraphics[width=\linewidth]{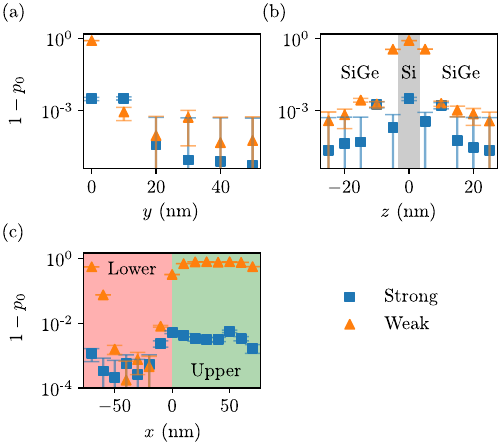}
    \caption{Occupation of the orbital excited states as $1-p_0$, after tracing out the spin state. Each data point corresponds to the result of a single defect being placed at the center of the channel with offsets in (a) the $y$-direction, (b) the $z$-direction, with respect to the center of the QW, and (c)~the $x$-direction. The center of the channel in the $x$-direction is defined as the mid-point between an upper and a lower clavier gate. The numerical uncertainty floor, resulting from limited computational resources, lies below $10^{-3}$. In the simulations we used a strong driving amplitude (blue squares) and a weak driving amplitude (orange triangles).}
    \label{fig:coeffs}
\end{figure}

The occupation of the orbital states throughout the evolution of the system provides valuable information about the strength of the interaction between the electron and the defect and the impact on the overall dephasing. This magnitude is also necessary to reconstruct the electron's density matrix in the Schrödinger frame (as opposed to the adiabatic formalism employed for the simulations) using the instantaneous eigenstate expansion~(\ref{eq:Schr}) and its density matrix equivalent in the Lindblad formalism. The occupation of the orbital state $\ket{l}$ is given by the diagonal elements of the density matrix $p_l = \rho_{l\uparrow,l\uparrow} + \rho_{l\downarrow,l\downarrow} = \text{Tr}\left[ \rho \kb{l(t)}{l(t)} \right]$. Throughout the evolution of the system, this quantity varies, but the largest variation occurs when the electron interacts with the defect. %
In Fig.~\ref{fig:coeffs} we show the minimum of the coefficient $p_0$---that is, the occupation of the orbital ground state at the point of maximal excitation---for different positions of the defect within the shuttling channel. Here we also show the effects of the strength of the electric confinement by comparing two different driving amplitudes: A strong driving amplitude $U_{i}^\text{AC} = 280$~mV for all gates S$_i$; and a weak driving amplitude $U_{i}^\text{AC} = 100$~mV. In Fig.~\ref{fig:coeffs}a, the orbital excitation is maximal when the defect is located at the center of the channel, with $y=0$\,nm, $z=0$\,nm and $x=35$~nm. This corresponds to a defect under the center of one of the upper clavier gates, as in Fig.~\ref{fig:potential}. %
For other positions of the defect along the $y$-direction, the probability of occupation of the excited orbital states decays exponentially as a function of the distance from the center of the channel. Similarly, Fig.~\ref{fig:coeffs}b shows that when the defect is kept at $y=0$\,nm while varying its vertical position either towards the gates ($z>0$\,nm) or towards the substrate ($z<0$\,nm), the occupation probability also decays exponentially, but with different rates, depending on whether the defect is inside the Si channel or in the SiGe part of the heterostructure. %
Most interestingly, when the defect is kept at the center of the shuttling channel $y=0$\,nm and $z=0$\,nm but at varying longitudinal position $x$, as in Fig.~\ref{fig:coeffs}c, we observe a strong dependence on the location of the clavier gates: A defect under an upper gate has a stronger influence on the orbital character of the envelope wave function than a defect under a lower clavier gate. %
Even though there is only one more oxide layer of $5\,$nm thickness differentiating the two clavier gate heights, this small difference changes the impact on the orbital excitation by more than three orders of magnitude, revealing unexpected levels of sensitivity. %
The impact of the defect is less pronounced when the confinement is stronger. A stronger confinement, either due to a larger potential or to a closer clavier gate, ensures a large orbital energy gap, thus reducing the chance of an orbital excitation when the electron meets the defect.

\begin{figure}
    \centering
    \includegraphics[width=\linewidth]{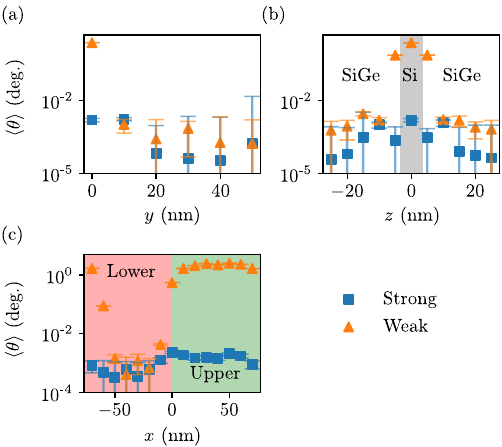}
    \caption{Expected dephasing after interaction with the defect and subsequent relaxation. Regardless of the position of the defect, the expected dephasing has always been evaluated at the same time $t=28$\,ns, long after interaction with the defect (electron in orbital ground state). Panels~(a-c) correspond to defect offsets along the $y$-, $z$- and $x$-direction, respectively, with $y=0$\,nm being the center of the QuBus, $x=0$\,nm the midpoint between two clavier gates and $z=0$\,nm the center of the QW. Here we have assumed a magnetic field of $0.5\,$T. Results are shown for both strong (blue squares) and weak (orange triangles) confinement, respectively.}
    \label{fig:phase}
\end{figure}

Due to $g$-factor differences between orbital levels, excited orbital state components will experience a modified phase evolution that differs from that of the ground state component. This difference starts at excitation---i.e., at the position of the defect---and is short lived due to the fast relaxation rate after interaction with the defect, i.e., when the orbital energy levels are split apart. Since the orbital relaxation process is non-deterministic, this accumulated phase difference will be a random quantity in a measurement. Nevertheless, the numerical treatment of this simulation using the Lindblad equation allows us to estimate the expected dephasing $\langle \theta \rangle$ after the evolution %
by evaluating the coherences, i.e, the off-diagonal elements of the density matrix. The expected dephasing $\langle \theta \rangle$ is then %
the phase of the ensemble-averaged spin-only coherence across all possible phonon-induced trajectories. We define this quantity as the argument of the coherences of the spin-only density matrix, after tracing out the orbital part
\begin{align}
    \langle \theta \rangle ={}&{} \text{Arg}\left( \langle \uparrow \vert\text{Tr}_\text{orb}\left[\rho\right] \vert\downarrow\rangle \right) \notag \\
    ={}&{} \text{Arg}\left(\text{Tr}_\text{orb} \left[\rho\right]_{\uparrow\downarrow}\right) \notag \\
    ={}&{} \text{Arg}\left(\rho_{s, \uparrow\downarrow}\right).
\end{align}
We evaluate this quantity at the end of the evolution, far from the defect and long after the system has reached orbital equilibrium.

Not surprisingly, Fig.~\ref{fig:phase} shows that the dephasing displays a similar behavior as the strength of the orbital excitation: A higher probability of excitation results in a longer occupation of an excited orbital and therefore in a larger accumulated phase. Similarly as in Fig.~\ref{fig:coeffs}, the dephasing is larger when the defect is in the middle of the channel, and the magnitude of the expected phase decreases as the defect is placed farther from the center of the channel. Interestingly, the magnitude of $\langle \theta \rangle$ is on the order of one degree in the case of weak confinement and a defect at the very center of the channel and it drops to below $10^{-3}$ degrees sufficiently far away from the shuttling lane, with a stronger driving amplitudes or when the defect is below an upper clavier gate. This should motivate the use of large driving amplitudes to mitigate the effects of decoherence due to charged defects in the channel, especially over long distances, as the probability of encountering a defect increases and the expected dephasing would also increase accordingly.

\begin{figure}
    \centering
    \includegraphics[width=\linewidth]{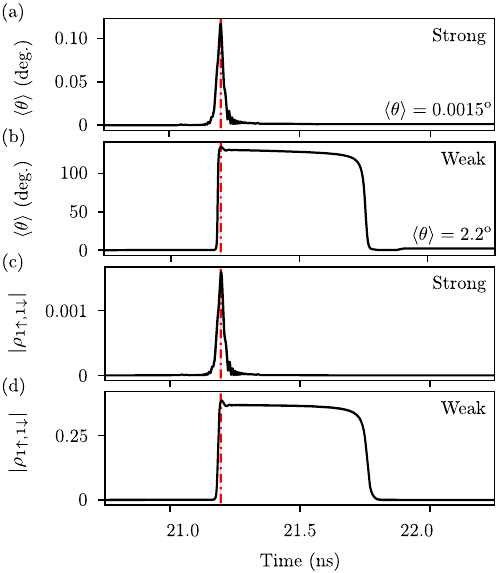}
    \caption{Expected dephasing as a function of time for (a)~a strong ($U_i^\text{AC} = 280$\,mV) and (b)~a weak driving amplitude ($U_i^\text{AC} = 100$\,mV). In both cases, the interaction with the defect takes place at the same point (vertical red line). Panels  (c) and (d) show the coherences $\rho_{1\uparrow, 1\downarrow}$ corresponding to the first excited orbital state for a strong and weak confinement, respectively. In the simulations we assumed a constant magnetic field $B=0.5\,$T.
    }
    \label{fig:time}
\end{figure}

More illustrative are the plots in Fig.~\ref{fig:time}a-b, where we show the expected dephasing at each point during the time evolution for a strong and a weak driving amplitude, respectively. In both cases the expected dephasing increases sharply at the position of the defect, signaling an orbital excitation, but the lower magnitude in the strong drive case of Fig.~\ref{fig:time}a compared to the weak drive case of Fig.~\ref{fig:time}b indicates a lower probability of excited state occupation in the first case. %
After the interaction with the defect, the system is in a superposition of orbital states, where the spin precession in the excited state evolves at a higher rate than in the ground state. This increases the expected dephasing $\langle \theta \rangle$. %

As soon as the orbitals are excited, the relaxation mechanism starts to take effect, decreasing the magnitude of the off-diagonal elements of the density matrix. This drives the system from a pure to a mixed state. In the strong confinement case, with $U_i^\text{AC} = 280$\,mV, the excitation is weak and the relaxation is very fast. During the short time where the excited orbitals are occupied, the coherences $\rho_{l\uparrow,l\downarrow}$ of the $l$-th excited orbital are also populated, as shown in Fig.~\ref{fig:time}c-d, and the phase of these quantities evolves according to the $g$-factor of the corresponding orbital, i.e., $\rho_{l\uparrow, l\downarrow} \sim \exp\left( -i \int \mathrm{d}t\, g_l\mu_\text{B} B/\hbar \right)$. The relaxation mechanism suppresses these coherences and at each time step, the collapse operators of the Lindblad equation contribute to the coherence of the ground orbital state $\rho_{0\uparrow,0\downarrow}$ by a small amount proportional to the excited orbital coherences. This results in an increase of the ground state expected dephasing.

Since the expected dephasing $\langle \theta \rangle$ is extracted by tracing out the orbital states, as the orbital states are populated, the expected dephasing increases. At the end of the evolution, when the system has relaxed to the orbital ground state again, only the coherence of the orbital ground state contributes to the total dephasing, hence the rise and fall of the expected dephasing as a function of time.

Due to the large orbital splitting and the lower probability of excitation in the strong drive case, the system relaxes quickly to the ground state, terminating the stochastic part of the phase evolution. In the case of weak confinement, on the other hand, the smaller energy splitting results in slower relaxation rates. This, combined with a larger occupation of the excited states, results in a larger phase accumulation that does not decay again until the phonon relaxation mechanism dominates the interaction.

\begin{figure}
    \centering
    \includegraphics[width=\linewidth]{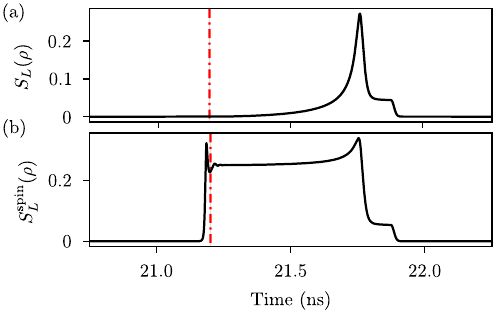}
    \caption{Purity of the quantum state as a function of time with a weak driving amplitude $U_i^\text{AC} = 100\,$mV. As a measure of purity, we evaluate (a)~the linear entropy of the density matrix and (b)~the linear entropy after tracing out the orbital part of the density matrix. The vertical red line marks the position of the defect. The simulations were carried out for a constant magnetic field $B=0.5\,$T.}
    \label{fig:purity}
\end{figure}

We evaluate the linear entropy $S_L (\rho) = 1-\text{Tr}[\rho^2]$ as a measure of purity, which vanishes for pure states and deviates from zero when the electron-phonon coupling results in a probabilistic mixing of spin and orbital states. From this measurement, we observe in Fig.~\ref{fig:purity}a a probabilistic mixing of the orbital states in the weak case, with a driving amplitude $U_i^\text{AC} = 100\,$mV. Interestingly, the entropy, in the weak driving case, rises some time after the first interaction with the defect. This indicates that the relaxation mechanism only dominates the electron-defect interaction around this point. The decay in entropy can be explained by the same mechanism that mixes the state: The relaxation mechanism brings the orbital part of the wave function to the ground state, which is a pure state, leaving only traces of the mixture in the spin component of the wave function.

Since we aim to encode quantum information in the spin of an electron, and any measurements will only capture the spin state of the electron, and not the orbital state, it is essential to evaluate the spin-fidelity, i.e., the purity of the electron density matrix after tracing out the orbital states. We thus evaluate the linear entropy $S_L^\text{spin} (\rho) = 1-\text{Tr}[\rho_s^2]$, with $\rho_s = \text{Tr}_\text{orb}[\rho]$. In Fig.~\ref{fig:purity}b we observe a rise in the linear entropy as soon as the defect induces orbital mixing. The entropy then plateaus until the onset of the relaxation mechanism.
The difference between the initial rise in linear entropy and the peak that occurs later in Fig.~\ref{fig:purity}b shows the extent of the defect interaction and the region with a slow relaxation mechanism. During this time, the phase continues to evolve at different rates for different orbital states.
From the moment the electron first interacts with the defect, the spin state is no longer pure, but after the system relaxes to the ground state and the coherences of higher orbitals have vanished, the electron transitions back towards a more pure state. After the interaction with the defect, some degree of mixture is still present in the spin density function of the electron, which results in a loss of coherence.

\section{Conclusion\label{sec:conclusions}}

We have presented a simulation framework for coherent spin shuttling in a QuBus device in the presence of localized charge defects. Our approach combines a realistic electrostatic model with time-dependent quantum dynamics governed by a Lindblad equation including a phonon-mediated orbital relaxation process. We solved the equations of motion in the adiabatic framework, reducing the computational cost of the problem. Our model enables us to capture both coherent evolution and defect-induced dephasing. This goes beyond previous numerical studies by explicitly including orbital degrees of freedom, phonon-assisted relaxation, and spin dephasing during transport.

Using this model, we quantified several figures of merit for qubit shuttling performance. In particular, we extracted the expected spin dephasing arising from electron–defect interactions and demonstrated its strong dependence on the drive amplitude and the gate geometry. Our results show that electrostatic disorder generated by proximal and sparse point-like charged defects might degrade shuttling performance depending on their volume density in SiGe. Clean SiGe heterostructures, low global magnetic fields and careful tuning of the driving potential amplitude provides an effective means to mitigate spin-dephasing during shuttling. 

For example, encountering a single negatively charged defect during electron shuttling induces an expected spin dephasing of approximately 2 degrees in the case of a weak driving amplitude $U_i^\text{AC} = 100$\, mV and a magnetic field of $B=0.5\,$ T (see Fig.~\ref{fig:phase}).
For such a small dephasing, this translates into a shuttling fidelity of $F\simeq \cos^2\left(\langle\theta\rangle/2\right)\simeq 0.9997$.
Assuming a volume density $n_d\approx 10^{14}$cm$^{-3}$ of charged defects at operation temperature 100\,mK, the probability of encountering such a charged defect over a shuttling distance of 10\,µm is only about 5\%, provided a point-like charged defect and an effective cross section of the envelope wave function of the electron spanning $\sim 60\,\text{nm}^2$, in line with the regions of large dephasing from Fig.~\ref{fig:phase}. This leads to an ensemble-averaged dephasing of only 0.1\,degrees per 10\,µm. Assuming also a uniform defect distribution and independent interactions, this expected dephasing scales linearly at short distances. In order to reach a more critical spin-dephasing of 5 or 10 degrees (where the fidelity would drop to $F\sim 0.99$), one would need to shuttle a distance of approximately 500\,µm or 1\,mm, respectively. If instead the defect density increases by a factor of 10, the probability of an encounter per 10\,µm rises to  $\sim 50\%$, and the expected dephasing becomes one degree over that same distance. Lowering the magnetic field by an order of magnitude could compensate for these effects. While this linear scaling provides useful intuition in the low-dephasing regime, care must be taken when extrapolating further as a nonlinear accumulation of phase errors can lead to the breakdown of this simplified picture.

Overall, the framework developed here provides a versatile tool for assessing the robustness of shuttling protocols against nanoscale disorder. Beyond the specific case of point-like charge defects, our methodology can be extended to study other sources of decoherence, such as valley excitations and valley-orbit coupling, spin-orbit coupling---which is a strong interaction in electron hole Ge/SiGe systems---, magnetic noise and other forms of electrostatic disorder, such as threading dislocations. Moreover, while simulations of coherent shuttling in a defect ensemble are computationally expensive, our simulation framework can shed light to these exploratory ideas by simulating the quantum evolution of a mixed state with different degrees of mixing.

As minimizing the defect density comes with many challenges and shuttling over large distances may be the unavoidable future of large-scale quantum computation, our results highlight the importance of operational optimization in addition to device design when engineering scalable spin-shuttling architectures. In this way, this tool offers both predictive power for experimental efforts and guidance for the design of next-generation shuttling devices that can meet the stringent requirements of large-scale silicon-based quantum computing.

Future work will expand the simulation framework to incorporate different types of sparse defects such as positively charged single-point defects, localized charge densities, and threading dislocations. Upcoming versions will also include valley physics, including inter-valley excitation and relaxation processes, and explicit spin-orbit interactions, which are relevant mechanism in Ge/SiGe spin-qubit devices. With these enhancements we aim to improve the predictive capabilities of the model and aid the design of more robust spin-qubit shuttle devices.

\begin{acknowledgments}
Simulations were performed with computing resources granted by RWTH Aachen University under projects thes1563 and rwth1681. This work was funded by the German Research Foundation (DFG) under Germany's Excellence Strategy - Cluster of Excellence ``Matter and Light for Quantum Computing'' (ML4Q) EXC 2004/2 - 390534769 and the Cluster of Excellence MATH+ (EXC-2046/1, 390685689).
\end{acknowledgments}

\appendix
\section{Electrostatic model~\label{app:comsol}}
We compute the electrostatic potential at the 2DEG layer using COMSOL Multiphysics\textsuperscript{\textregistered} with the \emph{Electrostatics (es)} interface in the stationary setting. Metal gates are treated as equipotential conductors (\textit{Terminal}). For the gate potential simulations, we impose lateral periodicity with $n_{\mathrm{cells}}=3$; for the defect geometry, we simulate $n_{\mathrm{cells}}=7$ without periodicity and extract the central three periods (two-period buffer on each side to avoid boundary effects). Exterior sidewalls use the charge-conservation (“zero charge”) condition. At material interfaces, we apply standard conditions: Continuity of the tangential electric field and, assuming the absence of surface charges, a continuous normal displacement field,
\( \mathbf n\times(\mathbf E_1-\mathbf E_2)=\mathbf 0 \) and \( \mathbf n\!\cdot(\mathbf D_1-\mathbf D_2)=0 \). Permittivities are set per material, as seen in table \ref{tab:mat_perms}. Geometry parameters of the simulated device are listed in Tab.~\ref{tab:bp_dims_single} and mesh parameters are given in Tab.~\ref{tab:mesh_unified}.

\begin{table*}[t]
\small
\begin{tabular}{p{0.5\linewidth} p{0.47\linewidth}}
\toprule
\textbf{Hardware parameter} & \textbf{Value (nm)} \\
\midrule
{\emph{Widths}} & \\
\addlinespace[2pt]
Lower metal gate width                    & 65 \\
Upper metal width                         & 65 \\
Inter-gate gap                            & 5 \\
Gate pitch                                & 70 \\
Total width of all cells                  & $n_{\mathrm{cells}}\cdot 280$ \\
\addlinespace[4pt]
{\emph{Heights}} & \\
\addlinespace[2pt]
Upper SiGe thickness                      & 30 \\
Lower SiGe thickness                      & 1000 \\
Si quantum well thickness                 & 7 \\
Si cap                                    & 1.5 \\
Gate oxide                                & 5 \\
Lower metal thickness                     & 27 \\
Upper metal thickness                     & 34 \\
Screening metal thickness                 & 10 \\
Heterostructure stack height              & 1038.5 \\
\addlinespace[4pt]
{\emph{Depths}} & \\
\addlinespace[2pt]
Screening region depth                    & 100 \\
Channel depth                             & 200 \\
Gate overhang past channel                & 20 \\
Total depth                               & 400 \\
\bottomrule
\end{tabular}
\caption{Blueprint geometry parameters used in all COMSOL runs, independent of boundary conditions. All widths/heights/depths are in nanometers (nm). The lateral span obeys \( w_{\mathrm{tot}}=n_{\mathrm{cells}}\cdot 280\,\text{nm} \), where a cell is a set of four clavier gates. For the \emph{gate potential}, we apply periodic boundary conditions with \( n_{\mathrm{cells}}=3 \) (three periods). For the \emph{defect} case, PBCs are not used; we simulate \( n_{\mathrm{cells}}=7 \) periods and extract fields from the central three periods, keeping two periods on each side as buffer. In both cases, outputs correspond to three periods of mesh at the QW level, which is sufficient to simulate shuttling across one full unit cell without any cropping.
}
\label{tab:bp_dims_single}
\end{table*}

\begin{table*}[t]
\centering
\begin{minipage}[t]{0.48\textwidth}
\vspace{0pt}
\centering
\small
\setlength{\tabcolsep}{3pt}
\renewcommand{\arraystretch}{1.15}
\begin{tabular}{p{0.59\columnwidth} p{0.195\columnwidth} p{0.195\columnwidth}}
\toprule
\textbf{Mesh parameter} & \textbf{Gates} & \textbf{Defect} \\
\midrule
Calibrate for                & General physics & General physics \\
Predefined setting           & Finer           & Custom          \\
Maximum element size         & 108             & 122             \\
Minimum element size         & 7.84            & 0.22            \\
Maximum element growth rate  & 1.4             & 1.3             \\
Curvature factor             & 0.4             & 0.2             \\
Resolution of narrow regions & 0.7             & 1               \\
\bottomrule
\end{tabular}
\caption{Finite-element mesh settings for the two simulation variants. A finer mesh is needed for capturing the features of the point-like defect accurately. Element sizes are in nanometers (nm).}
\label{tab:mesh_unified}

\end{minipage}\hfill
\begin{minipage}[t]{0.48\textwidth}
\vspace{0pt}
\centering
\small
\begin{tabular}{p{0.49\columnwidth} p{0.49\columnwidth}}
\toprule
\textbf{Material} & \textbf{Rel. permittivity} \\
\midrule
SiGe & 13.2\\
Si & 12\\
SiO$_2$ & 6\\
Metals & 1 \\
\bottomrule
\end{tabular}
\caption{Relative permittivities of the materials used.}
\label{tab:mat_perms}
\end{minipage}
\end{table*}

\section{Methods \label{app:methods}}

Our numerical analysis shows that the electron is well confined in one of the QDs, and the orbitals occupied during the evolution of the electron---even during the interaction with the defect---are always single-QD orbitals. This allows us to impose a cut-off of the Hilbert space and solve the dynamics of the electron within the adiabatic formalism.

We compute the potential using a spatial and temporal discretization that allows us to safely interpolate the potential between time steps when needed without introducing numerical artifacts. For the case at hand, with one point-like defect in the shuttling channel this implies a spatial resolution of $\Delta x = 0.5$\,nm and $\Delta y = 1$\,nm, and a temporal resolution $\Delta t = 0.05$\,ns, meaning that we generate a new set of potentials $V(t)$ with different gate voltages $U_i(t)$ for every 50\,ps of the time evolution.

As we solve the Lindblad equation in the adiabatic framework we need to ensure that the numerical scheme to solve the differential equation~(\ref{eq:lindblad}) converges. Using a second-order Magnus expansion, convergence is achieved when~\cite{blanes_magnus_2009}
\begin{align}
    \kappa_\text{ad} \equiv \Delta t \left| \frac{\bra{n(t)}\dot{H}(t) \ket{m(t)}}{E_n(t) - E_m(t)} \right | \ll 1,
\end{align}
for all states $\{n,m\}$, which imposes a condition on the length of the time step. 
Near the points where level crossings occur, shorter time steps are required. Additionally, due to the oscillating nature of the system in the adiabatic framework, we also impose a time step size that meets the von Neumann stability criterion $\kappa_\omega = \omega\Delta t \ll 2$, where $\hbar \omega$ is the largest relevant energy difference involved in the evolution of the system. This typically results in evaluations of the potential outside of the initial time grid, hence the need for time interpolation. Since the evaluation of the convergence criteria $\kappa_\text{ad}$ and the solution of the differential equation involve the evaluation of the time derivative of the Hamiltonian, we use a cubic spline interpolation to achieve a potential function that is smooth and has a smooth derivative.

We chose the $\kappa_\text{ad}$ and $\kappa_\omega$ that yield converged results in a reasonable amount of time. For most defect positions and both weak and strong driving amplitude, we enforce $\kappa_\text{ad} \le 0.01$ and $\kappa_\omega \le 0.05$. 
We monitor these quantities at each time step. If the inequalities are not satisfied, we shorten the duration of the time step recursively. Likewise, we increase the time step when possible to speed up the simulations. Simultaneously, at each time step we evaluate the relaxation rates using Eq.~(\ref{eq:fgr}).

Given a linear differential equation with a time dependent coefficient matrix
\begin{align}
    \frac{dy(t)}{dt} = A(t) y(t),
\end{align}
the formal solution reads
\begin{align}
    y(t) = \exp{\left(\Omega(t)\right)} y(0).
\end{align}
The generator $ \Omega(t) $ can be expanded in an infinite series involving a series of integrals of $A(t)$. The Magnus expansion \cite{Magnus1954} up to second order in the time step consists of truncating the expansion after the second term, resulting in 
\begin{align}
    \Omega(t) \simeq \int_{0}^{t} A(\tau) d\tau
            + \frac{1}{2} \int_{0}^{t} \int_{0}^{\tau_1} [A(\tau_1), A(\tau_2)] d\tau_2 \,d\tau_1.
\end{align}
When solving a Lindblad equation using this method, if the dissipative dynamics dominates over the coherent component of the time-evolution, numerical evaluation of the matrix exponential can become unstable in finite precision arithmetics. In order to avoid these numerical artifacts in the steps where the relaxation rates are very large, we use a Trotter-Suzuki approximation, where we separate the coherent operators and the decoherence terms and exponentiate them separately.

At each step of the evolution we keep track of the density matrix elements to evaluate the figures of merit presented in the main text.

%

\end{document}